\documentstyle[graphics,epsfig,floats,aps]{revtex}

\begin{document}

\draft
\catcode`\@=11 \catcode`\@=12
\twocolumn[\hsize\textwidth\columnwidth\hsize\csname@twocolumnfalse\endcsname
\title{Comments on $`$Rashba precession in quantum wire with
interaction'}
\author{Yue Yu}
\address{Institute of Theoretical Physics, Chinese Academy of
Sciences, P.O. Box 2735, Beijing 100080, China}

\date{\today}
\maketitle

\begin{abstract}

In a recent Rapid Communication (Phys. Rev. B {\bf 63}, 121210(R)
(2001)), Ha\"usler showed that the interaction between electrons
in quantum wires may enhance the persistent spin current arising
from Rashba spin-orbital coupling. In this Comments, we would like
to point out that this 'enhancement' comes from a misunderstanding
to the boosting persistent current in the Luttinger liquid theory.
A correct calculation will not give such an enhancement of the
persistent spin current. Meanwhile, we provide a Luttinger liquid
theory with Rashba spin-orbital interaction by bosonization, which
may show how the Rashba precession is in a Luttinger liquid.

\end{abstract}
]

In a recent Rapid Communication, Rashba precession in quantum
wires with interaction was discussed \cite{hau}. The author
explained that the enhancement of the Rashba effect in the
Shubnikov-de Hass measurement \cite{shu} may possibly caused by
the interaction between the electrons in quantum wires. A
bosonization form of the Luttinger liquid was used in order to
describe this enhancement. The bosonized Hamiltonian the author
obtained was (see eq. (5) in \cite{hau})
\begin{eqnarray}
H=\sum_{\nu=\rho,\sigma}\frac{\pi}{4L}(v_{\nu N}M_{\nu}^2+v_{\nu
J}J^2_{\nu})-m\alpha v_F J_\sigma+\sum_{q\ne 0}H_q.
\end{eqnarray}
Here, the author defined $M_{\nu}=M_{\nu R}+M_{\nu L}$ and $J_\nu=
M_{\nu R}-M_{\nu L}$. In Haldane's original paper, $M_{\nu R,L}$
are the particle numbers to be extra added to the ground state in
the right- or left-movers \cite{Hald}. They are integer. Now,
there is a contradiction, i. e., $J_\sigma$ is an integer while in
the ground state, $\delta H/\delta J_\sigma=0$ leads to
\begin{eqnarray}
J_{\sigma,0}=(2L/\pi)m\alpha v_F/v_{\sigma J}, \label{cen}
\end{eqnarray}
which is not an integer in general. On the other hand,
eq.(\ref{cen}) was the central result that the author of
\cite{hau} to claim the Rashba effect may be enhanced by
interaction, because of which, the Rashba length becomes short for
the repulsive interaction and Rashba effect is enhanced.

To solve the contradiction mentioned above and to see if the
Rashba length is really shorted, we thoroughly go through the
bosonization of the interaction electrons with Rashba effect in
quantum wires. Consider the free electrons  with Rashba
spin-orbital term on a quantum wire. The Hamiltonian in a second
quantization language reads
\begin{eqnarray}
H&=&\int dx \frac{1}{2m}\sum_{a=\pm
}\psi^*_a(x)[(-i\hbar\partial_x+aq_R)^2-q_R^2]
\psi_a(x)\nonumber\\
&=&\sum_k\epsilon_a(k)c^\dagger_{ka}c_{ka},
\end{eqnarray}
where $\psi_a=\sqrt{1/2}(\psi_\uparrow-ia\psi_\downarrow)$ for
$\psi_{\uparrow,\downarrow}$ being the electron fields with spin
$\uparrow$ and $\downarrow$; $c_{ka}$ is the Fourier component of
$\psi_a(x)$; $q_R=m\alpha$ is Rashba wave vector; and
$$\epsilon_a(k)
=\frac{1}{2m}[(k+aq_R)^2-q^2_R]$$ is the dispersion relation.
Linearizing the dispersion near the Fermi points $\pm k_F$, one
has
\begin{eqnarray}
H&\approx& H_R+H_L+E_0(M,J)=\sum_{k\sim
k_F}v_{Fa}(k-k_F)c_{ka}^\dagger c_{ka}\nonumber\\ &-&\sum_{k\sim
-k_F}v_{F\bar a}(k+k_F)c_{ka}^\dagger c_{ka}+E_0(M,J),
\end{eqnarray}
where $E_0(M,J)$ is the zero mode energy to be determined later;
$v_{Fa}=(k_F+aq_R)/m$ and $v_{F\bar a}=(k_F-aq_R)/m$. Define the
density operators in the momentum space
$$\rho_{qa}^{R,L}=\sum_{k\sim \pm k_F}c_{k+q,a}^\dagger c_{ka}$$
which obey commutation relations
\begin{eqnarray}
&&[\rho_{qa}^{R,L},\rho_{q'a'}^{\dagger
R,L}]=\frac{L}{2\pi}\delta_{a,a'} \delta{q,q'},\nonumber \\
&&[H^{R,L},\rho_{qa}^{R,L}]=\pm v_{F a,\bar
a}q\rho_{qa}^{R,L}.\label{comm}
\end{eqnarray}

Therefore, we can write down the bosonized Hamiltonian which
satisfying the commutation relation (\ref{comm})
\begin{eqnarray}
H_B=\sum_{q>0,a}v_{Fa}qb^\dagger_{qa}b_{qa} +\sum_{q>0,a}v_{F\bar
a}q\tilde b^\dagger_{qa}\tilde b_{qa}+E(M,J),
\end{eqnarray}
where $b_{qa}=\sqrt{2\pi/qL}\rho^R_{qa}$ and $\tilde
b_{qa}=\sqrt{2\pi/qL}\rho^{\dagger R}_{qa}$. Differing from a
well-known bosonized Luttinger liquid Hamiltonian, the left- and
right- mover Hamiltonians have  different sound velocities. Now,
we determine the zero mode energy. The ground state energy is
given by
\begin{eqnarray}
E_0=\int^{k_F}_{-k_F}dk\sum_an_a(k)\epsilon_a(k), \label{ge}
\end{eqnarray}
where $n_a(k)=1/2\pi$. The zero mode excitations include adding
extra particle to the ground state and boosting the Fermi sea by
$k\to k-\pi J_a/L$. That is, the energy increments are
\begin{eqnarray}
&& \delta_M E=\sum_a\biggl\{\int^{k_F+\pi M_a}_{-k_F-\pi M_a}
-\int^{k_F}_{-k_F}\biggr\}n_a(k)\epsilon_a(k),\nonumber\\
&&\delta_JE=\sum_a\biggl\{\int^{k_F-\pi J_a/L}_{-k_F-\pi J_a/L}
-\int^{k_F}_{-k_F}\biggr\}n_a(k)\epsilon_a(k). \label{inc}
\end{eqnarray}
It is easy  to see
\begin{eqnarray}
&&E_0(M,J)=\delta_M E+\delta_J E\nonumber\\
&&=v_F\sum_{\nu=\rho,\sigma}\frac{\pi}{4L}(M_{\nu}^2+J^2_{\nu})-q_R
v_F J_\sigma\nonumber\\
&&=v_F\sum_{\nu=\rho,\sigma}\frac{\pi}{4L}(M_{\nu}^2+\tilde
J^2_{\nu})-\frac{L}\pi v_F^2q_R^2,
\end{eqnarray}
where $M_\rho=\sum_aM_a$, $M_\sigma=\sum_aaM_a$ and so on; $\tilde
J_\rho=J_\rho$ and $\tilde J_\sigma=J_\sigma-(2L/\pi)q_Rv_F$.
According to Haldane \cite{Hald}, the periodic boundary condition
gives $(-1)^{M_\nu}=(-1)^{\tilde J_\nu}$. Namely, $M_\nu$ and
$\tilde J_\nu$ are integer and have the same odd-even.

A full Luttinger liquid Hamiltonian by adding the interaction
between electrons is given by
\begin{eqnarray}
H&=&H_B+H_I\nonumber\\
&=&\frac{1}2\sum_{q>0,a}[q(v_{Fa}+U_{qa})(b^\dagger_{qa}b_{qa}
+b_{qa}b^\dagger_{qa})\nonumber\\
&+&q(v_{F\bar a}+U_{qa})(\tilde b^\dagger_{qa}\tilde b_{qa}
+\tilde b_{qa}\tilde b^\dagger_{qa})\nonumber\\
&+&qV_{qa}(b_{qa}^\dagger\tilde b_{qa}^\dagger+\tilde
b_{qa}^\dagger b_{qa}^\dagger+b_{qa}\tilde b_{qa}+\tilde b_{qa}
b_{qa})]\nonumber\\
&+&\frac{\pi}{4L}\sum_\nu \{v_F(M_\nu^2+\tilde
J_\nu^2)+2[U_{0\nu}(M_{\nu,R}^2+M_{\nu L}^2)\nonumber\\
&+&2V_{0\nu}M_{\nu,R}M_{\nu,L}]\}+{\rm constant}.
\end{eqnarray}
Using the relation between the integers, i.e.,
$M_\nu=M_{\nu,R}+M_{\nu,L}$ and $\tilde
J_\nu=M_{\nu,R}-M_{\nu,L}$, one has
\begin{eqnarray}
H=\sum_{\nu=\rho,\sigma}\frac{\pi}{4L}(v_{\nu N}M_{\nu}^2+v_{\nu
J}J^2_{\nu})-q_R v_{\sigma J} J_\sigma+\sum_{q\ne
0}H_q,\label{ncen}
\end{eqnarray}
where $v_{\nu N}$ and $v_{\nu J}$ relate to $v_F$ by Haldane
controlling parameters determined by the interaction \cite{Hald}.
Differing from eq. (\ref{cen}), the persistent spin current in the
ground state determined by (\ref{ncen}) is simply
\begin{eqnarray}
J_{\sigma,0}=(2L/\pi)m\alpha, \label{ncurr}
\end{eqnarray}
 which is not renormalized by the
interaction.

To doubly check (\ref{ncurr}), we take another formalism of the
Luttinger liquid. The Luttinger liquid theory can be re-formalized
by using the exclusion statistics language \cite{wu}. In this
formalism, the Haldane controlling parameters, e.g.,
$e^{-2\varphi_\nu}$, are identified as the exclusion statistical
parameters, say $\lambda_\nu$. We consider the ground state energy
(\ref{ge}). Due to the $\lambda_\nu$-exclusion, the density
distribution of the fermi sea is given by
\begin{eqnarray}
n_\nu(k)=\frac{1}{2\pi \lambda_\nu} \label{nf}
\end{eqnarray}
 for $|k|<k_F$ but not simply
$\frac{1}{2\pi}$ \cite{wu}. Adding the extra particles corresponds
to enlarge the Fermi sea, i.e., in terms of (\ref{nf}), $\pm
k_F\to \pm(k_F+\pi\lambda_\nu M_\nu/L)$ . Boosting the Fermi sea
is still given by $k\to k-\pi J_\nu/L$. Therefore, similar to
(\ref{inc}), the increments of the energy due to the zero mode
excitations are given by
\begin{eqnarray}
\delta_M E&=&\frac{\pi}{4L}\sum_\nu(\lambda_\nu
v_F)M_\nu^2+...,\nonumber\\
\delta_J E&=&\frac{\pi}{4L}\sum_\nu(v_F/\lambda_\nu )J_\nu^2-q_R
(v_F/\lambda_\sigma) J_\sigma,
\end{eqnarray}
where $v_F\lambda_\nu\equiv v_{\nu N}$ and $v_F/\lambda_\nu\equiv
v_{\nu J}$; $'...'$ is proportional to $M_\nu$ and can be absorbed
into the re-definition of the chemical potential. Once more, we
obtain the persistent spin current (\ref{ncurr}) but not
(\ref{cen}).

In conclusions, we have shown that the interaction between
electrons does not renormalize the persistent spin current.
Therefore, the experimentally observed enhancement of the Rashba
effect is not related to the interaction in the way described by
Ref. \cite{hau}. However, the interaction does affect the Rashba
procession. It may be seen from $\sum_{q\ne 0} H_q$ in
(\ref{ncen}). Although the different sound velocities in the left-
and right-movers implies it is not an ordinary conformal field
theory with $c=1$ in each 'a'-sector, this $H_q$ may be
diagonalized. So the sound velocities $v_{Fa}$ and $v_{F\bar a}$
then the Rashba $q_R$ will be renormalized. Again due to the
technique of the ordinary conformal field theory can not be
directly applied, many details need to work out in order to see
the Rashba procession with interaction. These have gone beyond the
scope of this Comments and we will publish them in a separate
paper.

This work was supported in part by the NSF of China.


\begin{thebibliography}{99}
\bibitem{hau} W. Ha\"usler, Phys. Rev. B {\bf 63},121210(R)
(2001).
\bibitem{shu} J. Nitta, T. Akazaki, H. Takayanagi, and T. Enoki,
Phys. Rev. Lett. {\bf 78}, 1335 (1997). G. Engels, J. Lange, T.
Sch\"apers, and H. L\"uth, Phys. Rev. B {\bf 55}, R1958 (1997). T.
Sch\"apers, G. Engels, J. Lange, T. Klocke, M. Hollfelder, and H.
L\"uth, J. Appl. Phys. {\bf 83}, 4324 (1998). C.-M. Hu, J. Nitta,
T. Akazaki, H. Takayanagi, J. Osaka, P. Pfeffer, and W. Zawadzki,
Phys. Rev. B {\bf 60}, 7736 (1999). D. Grundler, Phys. Rev. Lett.
{\bf 84}, 6074 (2000). T. Matsuyama, R. K\"ursten, C.
Me$\beta$ner, and U. Merkt, Phys. Rev. B {\bf 61}, 15588 (2000).
\bibitem{Hald} D. M. Haldane, J. Phys. C {\bf 14}, 2585 (1981).
\bibitem{wu} Y. S. Wu and Yue Yu, Phys. Rev. Lett. {\bf 75}, 890
(1995). Y. S. Wu, Yue Yu and H. X. Yang, Nucl. Phys. B {\bf 604},
551 (2001).

\end{thebibliography}
\end{document}